\documentclass{emulateapj}

\shorttitle{Spatial Relationship between Flares and CMEs}
\shortauthors{Yashiro et~al.}

\begin{document}

\title{Spatial Relationship between Solar Flares and Coronal Mass
Ejections}

\author{S. Yashiro,\altaffilmark{1,2} G. Michalek,\altaffilmark{1,2,3}
S. Akiyama,\altaffilmark{1,2} N. Gopalswamy,\altaffilmark{2} and
R. A. Howard\altaffilmark{4}}

\altaffiltext{1}{Catholic University of America, Washington, DC 20064}

\altaffiltext{2}{NASA Goddard Space Flight Center, Greenbelt, MD 20771}

\altaffiltext{3}{Astronomical Observatory of Jagiellonian University,
Krakow, Poland}

\altaffiltext{4}{Naval Research Laboratory, Washington, DC 20375}

\begin{abstract}

We report on the spatial relationship between solar flares and coronal
mass ejections (CMEs) observed during 1996-2005 inclusive.  We
identified 496 flare-CME pairs considering limb flares (distance from
central meridian $\ge 45^\circ$) with soft X-ray flare size $\ge$ C3
level. The CMEs were detected by the Large Angle and Spectrometric
Coronagraph (LASCO) on board the {\it Solar and Heliospheric
Observatory} ({\it SOHO}). We investigated the flare positions with
respect to the CME span for the events with X-class, M-class, and
C-class flares separately. It is found that the most frequent flare
site is at the center of the CME span for all the three classes, but
that frequency is different for the different classes. Many X-class
flares often lie at the center of the associated CME, while C-class
flares widely spread to the outside of the CME span. The former is
different from previous studies, which concluded that no preferred
flare site exists. We compared our result with the previous studies
and conclude that the long-term LASCO observation enabled us to obtain
the detailed spatial relation between flares and CMEs. Our finding
calls for a closer flare-CME relationship and supports eruption models
typified by the CSHKP magnetic reconnection model.

\end{abstract}


\section{Introduction}

A solar flare is sudden flash of electromagnetic radiation (suggesting
plasma heating) in the solar atmosphere, and a coronal mass ejection
(CME) is an eruption of the atmospheric plasma into interplanetary
space. Both phenomena are thought to be different manifestations of
the same process which releases magnetic free energy stored in the
solar atmosphere. The spatial relation between flares and CMEs
contains information on the magnetic field configurations involved in
the eruptive process and hence is important for modeling them. Many
flare-CME models are based on the CSHKP (Carmichael, Sturrock,
Hirayama, Kopp \& Pneuman) magnetic reconnection model. The model
requires that a flare occurs just underneath of an erupting filament
which eventually becomes the core of the CME associated with the
flare. Normally the core corresponds to the center of the CME, thus
the CSHKP model requires that the flare occurs near the center of the
CME span.

Full-scale studies on the flare-CME relationship started in the 70s
and 80s with the CME observations obtained by the {\it Solwind}
coronagraph on board {\it P78-1} and the Coronagraph/Polarimeter
telescope on board the {\it Solar Maximum Mission} ({\it
SMM}). \citet{harri86} carried out a detailed analysis of three
flare-CME events observed by {\it SMM} and reported that flares
occurred near one foot of an X-ray arch, which is supposed to become a
CME. He also analyzed 48 flare-CME events observed by {\it SMM} and
{\it Solwind} and reported that many flares occurred near one leg of
the associated CMEs. This result, called the flare-ejection asymmetry,
is inconsistent with the CSHKP flare-CME model. \citet{kahle89}
examined 35 events observed by the {\it Solwind} and reported that
flare positions did not peak neither at the center nor at one leg of
the CMEs. They concurred with Harrison at the point that the
observations do not match with the CSHKP model, while disagreeing with
the result that flares are likely to occur at one leg of CMEs.  They
pointed out that the parameter employed by Harrison was biased, and
concluded that both observations are compatible with the fact that
there is no preferred flare site with respect to the CME span.  It
should be noted that the two studies applied different criteria for
the event selection. Harrison did not apply any criteria on flare
X-ray intensity, flare location, and CME span, while Kahler et
al. used only strong limb flares ($\ge$ M1 level; central meridian
distance (CMD) $\ge 40^\circ$) and wide CMEs (angular span $\ge
40^\circ$). Different criteria might produce different spatial
distributions, but the results in both the studies were inconsistent
with the schematic view of the CSHKP type flare-CME model.

The Large Angle and Spectrometric Coronagraph
\citep[LASCO;][]{bruec95} on board the {\it Solar and Heliospheric
Observatory} ({\it SOHO}) mission has observed more than 11,000 CMEs
from 1996, which provides a great opportunity to investigate the flare-CME
relationship. \citet{harri06} reviewed several flare-CME studies and
stated that "the pre-SOHO conclusions about relative flare-CME
locations and asymmetry are consistent with many recent studies."
However, systematic statistical study is needed before reaching a firm
conclusion. In this paper we revisit this issue using the large CME
data obtained by {\it SOHO} LASCO.

\begin{figure*} 
\epsscale{0.90} \plotone{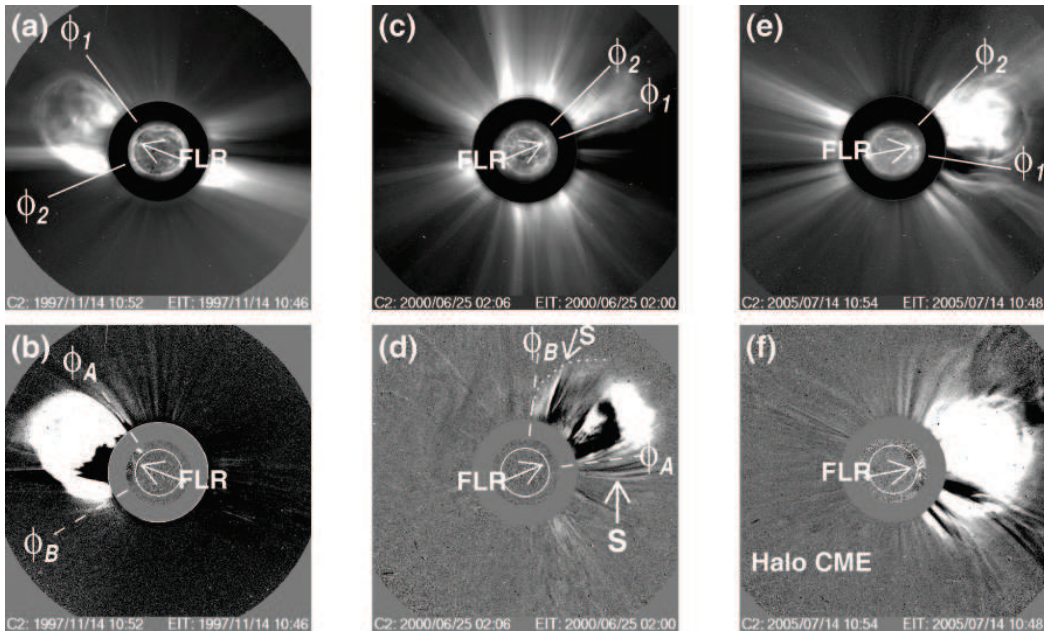} 
\caption{Three CMEs observed by {\it SOHO} LASCO to illustrate the
measurement of CME span. The top row shows direct images used to
measure the main CME body, and the bottom row shows corresponding
running difference images used to measure the whole CMEs. $\phi_1$ and
$\phi_2$ indicate the PAs of side edges of the main CME body, and
$\phi_A$ and $\phi_B$ indicate those of the whole CME. Arrows point to
the position of the flares associated with the CMEs.}
\end{figure*}

\section{Data and Analysis}

Solar flares are continuously monitored by the X-ray Sensor (XRS) on
board the {\it Geostationary Operational Environmental Satellite}
({\it GOES}) missions. The XRS observes the whole-sun X-ray flux in
the 0.1-0.8 nm wavelength band to detect solar flares. The flare
location has been determined by H$\alpha$ images obtained by
ground-base observatories and X-ray images obtained by Soft X-ray
Imager (SXI) on {\it GOES}. All flares have been listed in the Solar
Geophysical Data (SGD) and the online solar event
report\footnote{\url{http://www.sec.noaa.gov/ftpmenu/indices.html}}
compiled by NOAA Space Environment Center. From the online report we
selected limb flares (CMD $\ge 45^\circ$) with soft X-ray flare size
$\ge$ C3 level.

We used the {\it SOHO} LASCO CME
Catalog\footnote{\url{http://cdaw.gsfc.nasa.gov/CME\_list/index.html}}
\citep{yashi04} to investigate the CME associations. The CME
candidates associated with a given flare were searched within a 3 hr
time window (90 minutes before and 90 minutes after the onset of the
flare).  However, because the time window analysis by itself could
produce false flare-CME pairs, we checked the consistency of the
associations by viewing both flare and CME movies in the Catalog. We
played movies obtained by the Extreme ultraviolet Imaging Telescope
(EIT) on {\it SOHO} and Soft X-ray Telescope (SXT) on {\it Yohkoh} to
look for any eruptive surface activities (e.g., filament eruptions,
dimmings, and arcade formations) associated with the flares. All
flares can be divided into those with and without CMEs except for some
in which the eruptive signatures were obscure. We excluded such
uncertain flare-CME pairs from this analysis. From 1996 to 2005, we
found 496 definitive flare-CME pairs.

A typical CME consists of a bright frontal structure (leading edge),
followed by a dark cavity, and a bright core. This configuration is
called the CME three-part structure \citep{illin85,webb88}.  The
bright core corresponds to the erupting filament
\citep{webb87,gopal03a}. There is an issue whether narrow CMEs have
the three-part structure or not \citep[e.g.][]{gilbe01}, but at least
for large CMEs, this structure is fundamental. In this paper we refer
to the three-part structure as the main CME body. Some CMEs possess a
faint envelope outside of the main CME body. The envelope might be a
shock wave driven by the CME \citep{sheel00,vourl03,ciara05}, thus
there is a problem whether the envelope is a part of the CMEs or
not \citep[see][]{stcyr05}. However, since there is no established way
to identify a shock by coronagraph observation itself, we have
included the envelope structures as a part of the CME and refer to all
the CME features as the whole CME.

For the comparison between flare and CME positions, it is ideal if we
could measure the position angles\footnote{PA is measured
counterclockwise from Solar North in degrees} (PAs) of the CME edges
on the solar limb. The innermost coronagraph C1 is the best, but its
data are not available for the most of the CMEs. Since several CMEs
did show non-radial motion \citep{gopal00,zhang04}, we measured PAs of
the CME edges in C2 images as close to the occulting disk as possible.

\begin{figure*} 
\epsscale{1.0}\plotone{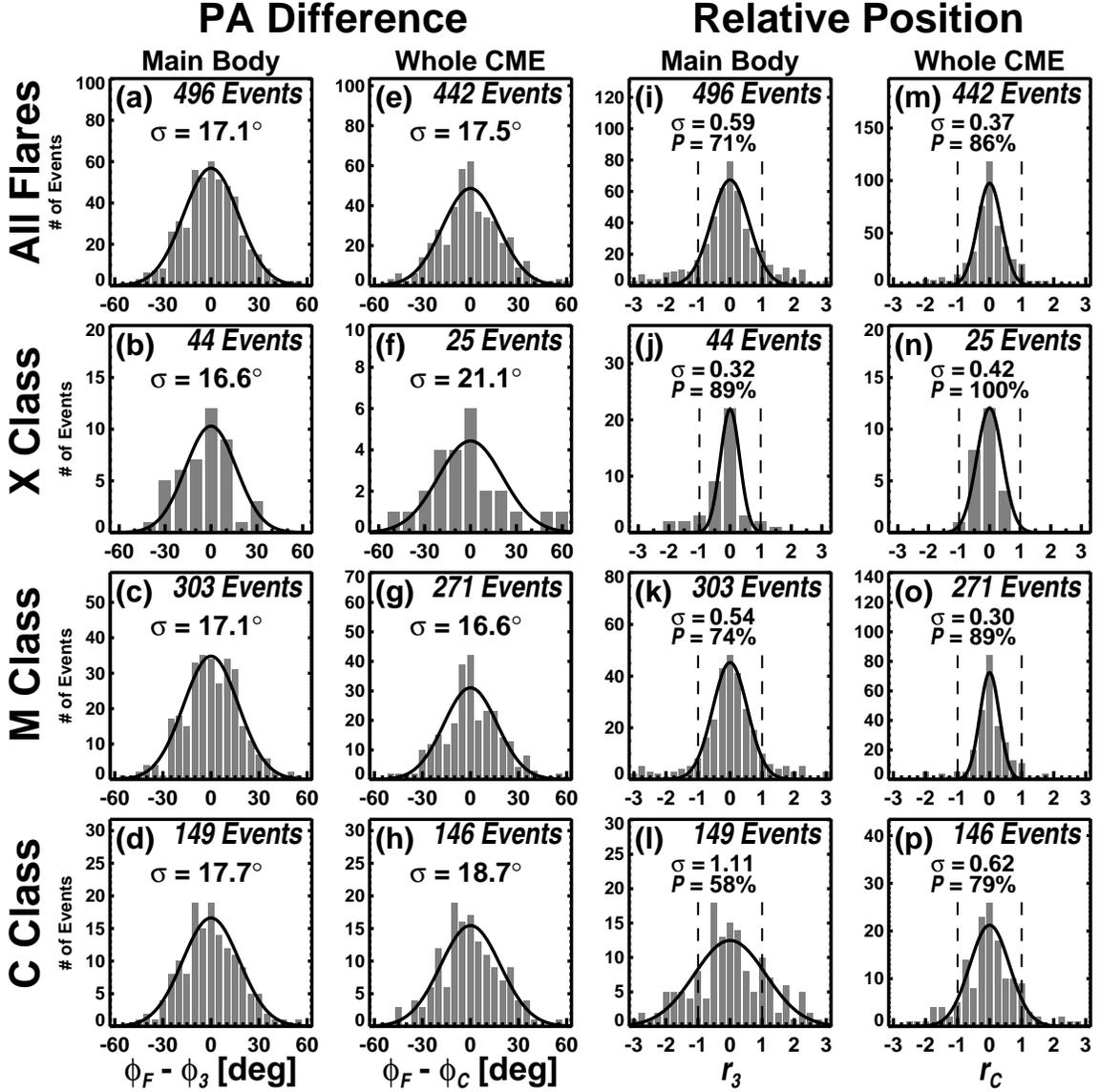} 

\caption{
Distributions of flare positions with respect to the CPA of the
CME. The first and second columns show the PA difference in degree for
the main CME body and for the whole CME, respectively. The standard
deviation ($\sigma$) obtained by Gaussian fit is shown in each plot. The
third and forth columns show the distributions of PA differences
normalized by the half CME span. The vertical dashed lines mark the
two side edges of the CMEs. $P$ is the percentage of the flares lying
inside of the CME span. The second, third, and forth rows correspond
to the events with X-class, M-class, and C-class flares, respectively.}

\end{figure*}

Figure~1 illustrates how we measured the PAs of the main CME body and
the whole CME.  Top panels are LASCO C2 images for three CMEs with the
corresponding running difference images (previous images are
subtracted to enhance the faint structure of the CMEs) in the bottom
panels. The side edges of the main CME body (the whole CME) are
denoted by $\phi_1$ and $\phi_2$ ($\phi_A$ and $\phi_B$). The CME on
1997 November 14 (Figs.~1a and 1b) did not have an envelope, thus the
$\phi_1$ ($\phi_2$) and $\phi_A$ ($\phi_B$) are identical. On the
other hand, the CME on 2000 June 25 had a faint envelope to the north
of the main CME body (Figs.~1c and 1d). Since it is hard to see it in
print, we traced out the edge of the envelope by a dotted curve on
Fig.~1d. The northern edge of the envelope denoted by $\phi_B$ is used
for the edge of the whole CME. The black-and-white radial features
(denoted by $S$) at both sides of the CME are a signature of streamer
shift caused by the expansion of the CME. We should note that we did
not use them for the determination of the edges of the whole
CME. Since we can not see an envelope to the south of the CME, the
southern edges of the main CME body and whole CME ($\phi_1$ and
$\phi_A$) are almost identical. The CME on 2005 July 14 appeared in
the C2 FOV at 10:54 UT (Figs.~1e and 1f). The CME had a clear
three-part structure with a faint envelope. The envelope covered the
occulting disk at 11:54 UT, thus the CME is listed as a halo
\citep{howar82} in the CME catalog. In this case $\phi_A$ and $\phi_B$
cannot be defined.

The location of a CME is represented by the central position angle
(CPA), which is defined as the mid-angle of the two side edges of the
CME in the sky plane. We define the CPA of the main CME body as
$\phi_3=(\phi_1+\phi_2)/2$ and that of the whole CME as
$\phi_C=(\phi_A+\phi_B)/2$. The PAs of flares ($\phi_F$) are computed
from their location in heliographic coordinates listed in NOAA SGD.
The angular span of the main CME body ($\omega_3$) and the whole CME
($\omega_C$) is defined as the difference between the two side edges
[$\omega_3=\phi_2-\phi_1$; $\omega_C=\phi_B-\phi_A$].

\section{Results}

\subsection{PA Difference}

The distributions of differences between flare PAs and CME CPAs for
the main CME body and for the whole CME are shown in Figures~2a and
2e, respectively. Fifty four halo CMEs are not used in Figure 2e since
their $\phi_C$ cannot be defined. Both the distributions are very
similar and are well represented by Gaussians. The standard deviation
is 17.1$^\circ$ for the difference between the flares and main CME
bodies and 17.5$^\circ$ for the difference between the flares and
whole CMEs. The average (median) angular span is 52.2$^\circ$
(43.8$^\circ$) for the main body of CMEs and 89.5$^\circ$
(75.0$^\circ$) for the whole CMEs. Thus the PA differences between
flares and CMEs are smaller than the angular span of the CMEs. One
might think that the flare site is inherently close to the center of
the CME and the non-radial motion below the C2 occulting disk produces
the PA differences. In order to explain a difference of 17$^\circ$, a
CME needs to erupt 30$^\circ$ away from the radial direction.

We separated the events into three groups according to their flare
intensity and made the same plots for each group. The second, third,
and forth rows in Figure~2 correspond to the events with X-class,
M-class, and C-class flares, respectively. The standard deviation is
shown in each plot, which ranges from 16.6$^\circ$-17.7$^\circ$ for
the main CME body and 16.6$^\circ$-21.1$^\circ$ for the whole CME. The
distributions of PA differences in the three groups are almost
identical, suggesting that the events with weak flares (below C3
level) have a similar distribution.

\subsection{Relative PA Difference}

In order to investigate the flare position with respect to the main
CME body (frontal structure), we normalize the PA differences by the
half angular span. We define relative flare location
$r_3=(\phi_F-\phi_3)/0.5\omega_3$. The $r_3=\pm1$ indicates the flare
is located at either leg of the CME frontal structure. The $r_3=0$
indicates that the flare is located at the center of the CME span. The
distribution of $r_3$ is shown in Figure~2i. It is clear that most of
the flares are located under the span of the main CME body. Out of the
496 flares, 350 (or 71\%) resided under the span of the main CME
body. Figure~2m is the same as the Figure~2i, but for flare locations
with respect to the edges of the whole CMEs
[$r_C=(\phi_F-\phi_C)/0.5\omega_C$].  Again, we exclude 54 full halos
($\omega_C=360^\circ$). Out of the 442 flares, 379 (or 86\%) resided
under the angular span of the whole CMEs. Both the distributions are
well represented by Gaussians with standard deviations of 0.59 for the
main CME body and 0.37 for the whole CME. In both the cases, the peak
of the Gaussian is around zero, meaning that the flares frequently
occur under the center of the CME span, not near one leg (outer edge)
of the CMEs.

As we did for the PA difference distributions, we separated the events
into three groups according to their flare intensity. The second,
third, and forth rows in Figure~2 correspond to events with X-class,
M-class, and C-class flares, respectively. $\sigma$ is the standard
deviation and $P$ is the percentage of the flares occurring under the
CME span. We found that all distributions have a peak around zero,
while the width of the distributions is different for different flare
levels. The flare-CME events with X-class flares (hereafter X-class
events\footnote{Similarly we labeled flare-CME events with M-class
(C-class) flares as M-class (C-class) events.}) have a narrower
distribution suggesting that many X-class flares lie under the center
of the CME span. On the other hand, the C-class events have a broader
distribution and a significant number of events occurred outside of
the CME span.

We do not see a significant distinction in the PA difference
distributions among the three flare levels, but we do see a difference
in the relative position distributions.  Since the relative position
is defined by the PA difference normalized by the CME half span, the
distinction shown in Figure 2 results from the difference in CME
span. The average angular span of the main CME body (the whole CME) is
87$^\circ$ (224$^\circ$), 54$^\circ$ (124$^\circ$), and 38$^\circ$
(75$^\circ$) for X-, M-, and C-class events, respectively. As reported
by the previous studies \citep[e.g.,][]{yashi05}, by means of
statistics, CMEs associated with stronger flares have larger angular
span. 

\begin{figure}[t] \epsscale{1.00}\plotone{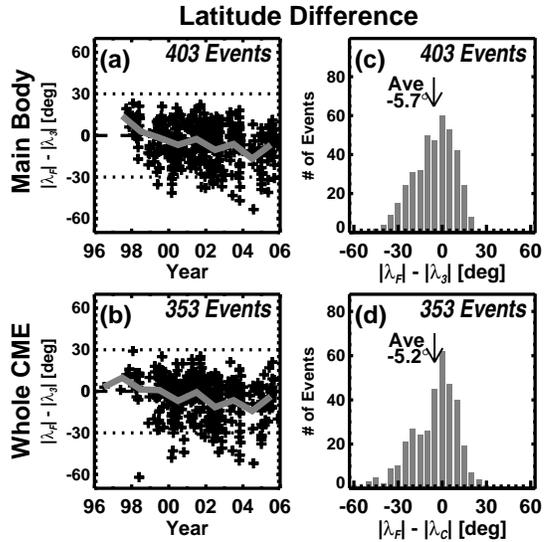} 
\caption{Solar cycle variation ({\it left}) and distribution ({\it
right}) of the latitude difference between flares and
CMEs. $\lambda_F$, $\lambda_3$, and $\lambda_C$ are latitudes of the
flares, main CME body, and whole CME, respectively. The flare-CME
pairs occuring in different hemispheres (e.g. a flare in the northern
hemisphere and a CME in the southern) were excluded.}
\end{figure}

\subsection{Latitude Difference}

Even though the flare PAs tend to be close to those of CMEs, we need
to point out some systematic offsets between flares and CMEs stemming
from the varying influence of the global solar magnetic field as a
function of the solar cycle. \citet{gopal03a} examined the
relationship between prominence eruptions (PEs) and CMEs and found
that, during solar minimum, the CPA of CMEs tend to be closer to the
equator compared to those of PEs, while no such effect was seen during
solar maximum. A similar relation is expected between flares and
CMEs. In order to examine whether flare positions are equatorward or
polarward with respect to the CMEs, we have shown their latitudinal
differences in Figures~3.  The CME latitudes ($\lambda_3$ and
$\lambda_C$) were calculated from CPAs of the main CME body ($\phi_3$)
and whole CME ($\phi_C$), respectively. The flare-CME pairs occurring
in different hemispheres were excluded.

The gray lines in Figs~3a and 3b show annual average of the latitude
difference.  In spite of the small data set during solar minimum, we
see a positive offset in 1997, meaning that CMEs occurred in lower
latitudes as compared to flares. This is consistent with the result of
Gopalswamy et al. On the other hand, during solar maximum, we see a
negative offset, indicating that flares occurred in lower latitudes as
compared to CMEs.  This is different from Gopalswamy et al. who
reported that no systematic offset exists between PE and CME
latitudes. The difference may result from the exclusion of
high-latitude flares from our analysis.  Since we did not include
small flares (below C3 level), the high-latitude flares (e.g. X-ray
arcade formations associated with the eruption of polar crown
filaments) were excluded.  Therefore the sampled flares mainly
occurred in active regions, i.e. in low latitudes. We should note that
high-latitude CMEs are not associated with active-region flares, but
appear frequently during solar maximum \citep{gopal03b}. During the
declining phase of solar cycle 23, CMEs have gradually started
clustering around the equator, while active regions have remained
around the equator. This is why we still see the negative offset in
2004 and 2005. The positive offset between flare and CME latitude is
likely to resume with the start of solar cycle 24.

One would think that the existence of the latitude offset is
inconsistent with the result that flares frequently occur under the
center of the CME span. In order to check this, we have shown the
distributions of latitude differences in figures~3c and 3d. Because of
the exclusion of the flare-CME events from different hemispheres, the
distributions became narrower since the excluded events have
relatively larger PA differences. We can see that both the
distributions are asymmetric with a broad tail on the left. However,
the most frerqent bin stays as 0$^\circ$ and the average (median)
difference is -5.7$^\circ$ (-3.9$^\circ$) for the main CME body and
-5.2$^\circ$ (-2.0$^\circ$) for the whole CME.  We conclude that there
are systematic offsets between flares and CMEs, but such an offset is
only a small fraction ($\sim 10\%$) of the CME span.

\section{Summary \& Discussion}

We investigated the spatial relationship between solar flares and CMEs
for 496 pairs occurring from 1996 to 2005. It is found that the
distribution of the difference between flare PA and CME CPA can be
represented by Gaussian centered at zero with a standard deviation of
$\sim17^\circ$, and the distribution does not change with the flare
level. We examined the flare positions with respect to the CME span
and found that the most probable flare site is the center of the CME
span for all flare levels, but the width of the distributions is
different for different flare levels. For C-class events, the flare
positions widely scattered with respect to the CME span, while for
X-class events, most of the flares lie under the center of the CME
span. The result is suitable for flare-CME models typified by the
CSHKP reconnection model. 

\begin{figure}[bt] 
\epsscale{1.0}\plotone{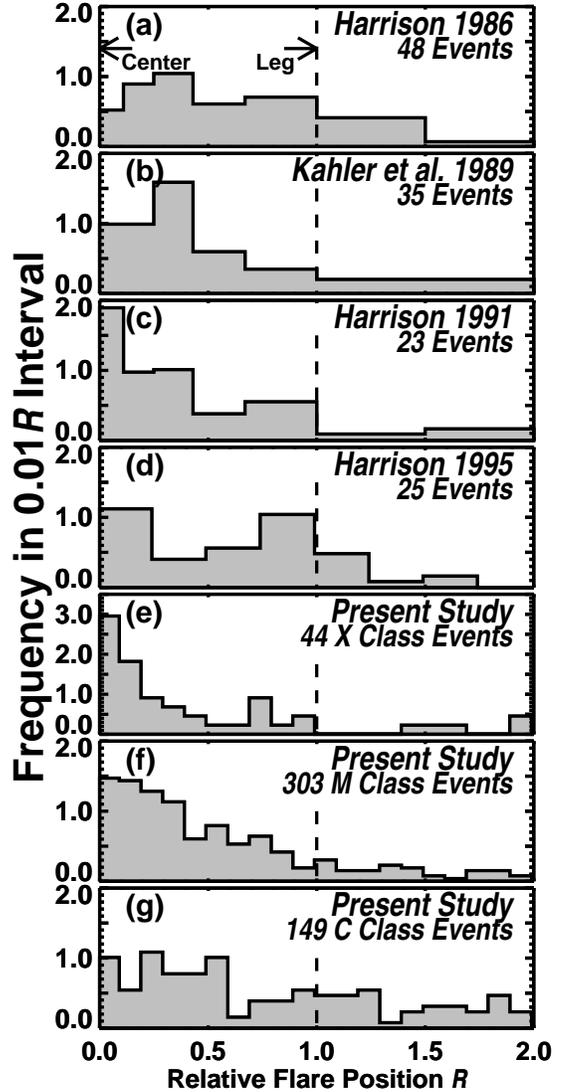} 
\caption{
Flare position ($R$) with respect to the CME span. $R=0$ for a flare
centered under the CME, $R=1$ for a flare at one leg of the CME, and
$R>1$ for a flare lying outside of the CME span.}
\end{figure}

\begin{deluxetable*}{lcclc}
\tabletypesize{\scriptsize} 
\tablecolumns{5} 
\tablewidth{0pc} 
\tablecaption{Summary of studies on the spatial relationship between
flares and CMEs}
\tablehead{ 
\colhead{} & \colhead{\# of Events}   & \colhead{Satellite}&
\colhead{Data Period} & \colhead{Remarks}}
\startdata 
Harrison 1986 (H86)      & 48 & {\it Solwind} and {\it SMM} & $1979-82$ & \nodata \\
Kahler et al. 1989 (K89) & 35 & {\it Solwind}               & $1979-82$ & $I_P$\tablenotemark{1} $\ge$ M1; CMD\tablenotemark{2} $\ge 45^\circ$; WD\tablenotemark{3} $\ge 40^\circ$ \\
Harrison 1991 (H91)      & 23 & {\it SMM}                   & $1984-87$ & \nodata \\
Harrison 1995 (H95)      & 25 & {\it SMM}                   & $1986-87$ & CMD $\ge 40^\circ$ \\
Present Study           & 496 & {\it SOHO}                  & $1996-05$ & $I_P \ge$ C3; CMD $\ge 45^\circ$ \\

\enddata 
\tablenotetext{1}{Peak X-ray Intensity of a flare}
\tablenotetext{2}{Central Meridian Distance of a flare}
\tablenotetext{3}{Angular Span of a CME}
\end{deluxetable*} 



\subsection{Comparison with Previous Studies}

Let us compare our results with four previous studies,
\citet[hereafter H86]{harri86}, \citet[hereafter K89]{kahle89},
\citet[hereafter H91]{harri91}, and \citet[hereafter
H95]{harri95}. All previous studies examined flare positions with
respect to the CME span using data obtained by {\it Solwind} or {\it
SMM}. Since the observational capability of the pre-SOHO coronagraphs
is thought to be lower, it is possible that they did not detect the
faint envelope around the three-part CME structure. Therefore, for the
proper comparison, we used $r_3$, the flare positions with respect to
the main CME body.

H86 employed the parameter $\alpha=(\phi_2-\phi_F)/(\phi_F-\phi_1)$
(In Figure~4 of H86, $\phi_2$, $\phi_F$, and $\phi_1$ correspond to
letters $A$, $B$, and $C$). He found a significant peak at
$\alpha=0-0.19$, meaning that flares often occur at the one leg of the
CMEs. However, K89 pointed out that the bin size in equal $\alpha$ is
biased; the smallest bin ($\alpha=0-0.19$) is about 3 times more
probable than the largest bin ($\alpha=0.8-1.0$) for the random
distribution of flare positions. They employed the parameter
$R=|\phi_F-\phi_3|/0.5\omega_3$, which is the absolute of $r_3$. The
relation between parameters $\alpha$ and $R$ is
$R=(1-\alpha)/(1+\alpha)$.

In order to compare the previous studies properly, we converted the
distributions in equal $\alpha$ into those in equal $R$. For each
$\alpha$ bin, the frequency ($f$) in 0.01 $R$ intervals in percentage
is computed by dividing the number of events ($n$) by both the total
number of events ($N$) and the interval of $R$ ($dR$),
i.e. $f=n/N/dR$. For example, H86 examined 48 flare-CME events and
found that 3 of them are in the bin of $\alpha=0.8-1.0$, which
corresponds to $R=0-0.11$ ($N=48$, $n=3$, and $dR=0.12$). Then we
obtained the frequency for $R=0-0.11$ bin to be 0.52 (=3/48/0.12). We
carried out the same conversion for other bins. However, treatment of
the $\alpha=0-0.19$ bin is not easy. Harrison used 30 {\it Solwind}
events compiled by Sheeley et al. (1984) and 18 {\it SMM} events
compiled by Sawyer. We examined the Sheeley et al.'s list and found 7
events lying outside the CME span. Such flares should have negative
$\alpha$, but there is no corresponding bin in Fig.~5 of H86. Thus we
supposed that the 7 events were included in $\alpha=0-0.19$ bin, and
determined their $R$ using the Sheeley et al's list. Unfortunately we
could not locate Sawyer's 18 events. Thus we assumed that the same
fraction of the events in the bin inherently lied outside of the CME
span. The result of this analysis is shown in Figure~4a. By the
conversion of equal $\alpha$ to equal $R$ distribution and the special
treatment of $\alpha=0-0.19$ bin, the peak at the $\alpha=0-0.19$ bin
in Fig.~5 of H86 disappeared. For K89, we obtained the $R$
distribution from their Table~3. They reported 7 out of the 35 flares
occurring outside the CME span. However we could not find out how far
apart the flares were located from the nearest CME leg.  Thus we
assumed that the 7 events resided in $R=1.01-2.00$. The $R$
distribution of K89 is shown in Figure~4b. H91 has a histogram of the
$\alpha$ distribution and their conversion to the $R$ distribution was
straightforward (Fig.~4c). H95 does not have a histogram of spatial
distribution, but has the scatter plots of $R$ vs. flare intensity and
$R$ vs. flare duration. We read the $R$ value from the plots and made
a histogram of the $R$ distribution (Fig.~4d). For the present study,
we plotted the $R$ distribution for the X-class, M-class, and C-class
events in Figure~4e, 4f, and 4g, respectively.

As we showed in Section~3.2, the $R$ distribution varies according to
the scale of the flare-CME events. Therefore, in the comparison with
previous studies, we should pay attention to their data source and
selection criteria, which are summarized in Table~1. The second and
third columns show that the number of events and satellites used in
each study. The forth column shows the study period. H86 and K89 used
data during solar maximum, while H91 and H95 used data during solar
minimum. The present study covers the almost whole of the solar cycle
23, but many events were obtained during solar maximum. The last
column is for event selection criteria in each study. K89 and the
present study used only strong flares ($\ge$ M1 and $\ge$ C3,
respectively), but other studies by Harrison did not eliminate weak
flares. K89, H95, and the present study used limb events only, which
reduces the projection effects.

Except for H95 (Fig.~4d), the three previous studies show a trend that
the flare occurred near the center of CMEs rather than at the
edges. The majority of the events in H95 were C-class flares, thus we
should compare the result of H95 with our C-class events
(Fig.~4g). The C-class events in our data show a trend that flare
occurred near the CME center, but the trend is very weak. Therefore it
is not surprising that the examination of the 25 events can not see
such weak trend. On the other hand, H91 shows a strong peak at the
center of the CME even though the data were obtained during solar
minimum (1984-87).  We could not find a statement of exclusion of disk
events, thus the projection effects might produce the peak (Apparent
CME span becomes larger than inherent span if the distance from the
limb is farther). Except for the lack of flares under the center of
the CME span ($R<0.25$), the $R$ distribution of K89 (Fig.~4b) is
similar to that of our M-class events (Fig.~4f), which can be
explained by their selection criterion about flare intensity ($\ge$ M1
level). By the same token, because 30\% and 57\% of the H86 events
were X-class and M-class flares, respectively, Fig.~4a should be the
similar to Fig.~4e or Fig.~4f. However the distribution is similar to
that of C-class events (Fig.~4g). Additionally H86 also lacks flares
under the center of the CME span ($R<0.25$). The average CME span of
H86 events (64$^\circ$) is larger than that of our M-class events
(56$^\circ$), suggesting that the different capability between SOHO
LASCO and previous coronagraphs does not explain the H86 lack of
flares under the center of the CME span. Except for this discrepancy,
all the flare position distributions reported in the previous studies
are consistent with our results. The long-term LASCO observation
enabled us obtain a large number of flare-CME pairs from small to
large events for the first time that revealed the detailed spatial
relation between flares and CMEs.

\subsection{Flare-CME Geometry}

The flare-CME asymmetry found by H86 has been the basis of the claim
that flare-CME observations are inconsistent with the schematic
picture of the CSHKP type flare-CME models. In this paper we show that
most of the X-class flares are located at the center of the CME span
while a significant number of C-class flares reside near the edge or
even outside of the CME span. The CSHKP type flare-CME models are well
suited for strong events, but may not be applicable for the many weak
events.  The other extreme is the non-eruptive (or compact) flares
which do not involve any mass motion, and hence their geometry may not
be appropriate for CSHKP models.

The flare-CME geometry is possibly different between weak (narrow) and
strong (wide) events. This is related to the issue whether narrow CMEs
are physically distinct from general CMEs
\citep{kahle89,kahle01}. \citet{reames02} presented two types of
flare-CME geometry which are responsible for two types of solar
energetic particle (SEP) events, i.e. impulsive and gradual
events. The gradual SEP events are associated with large CMEs, which
fit CSHKP type models, while the impulsive SEPs are associated with
narrow CMEs that fit the X-ray jet model
\citep{shimo00}. \citet{bempo05} reported that blob-like narrow CMEs
in a streamer (called "streamer puffs") differ from general CMEs
\citep[see also][]{moore07}. The streamer puffs are associated with
weak flares (below C4 level) and their schematic picture clearly
explains the flare-CME asymmetry.

Even X-class events, some of them showed the clear flare-CME
asymmetry. A good example is the event on 2002 May 20. The X2.1 flare
at 15:21 UT located at one edge of the CME at 15:50 UT. Another
example is the event on 2003 November 3 (The X2.7 flare at 01:09 UT
and the CME at 01:59 UT). In both cases EIT dimmings were clearly
observed only on the CME side of the flares. It is important to
investigate the flare-dimming asymmetry for understanding the origin
of such flare-CME asymmetry.

We examined the flare positions with respect to CME spans using LASCO
data and found that most frequent flare site is the center of the CME
span. However, since we examined the spatial relationship using limb
events, our finding can apply only in the latitudinal direction. The
flare-CME geometry in the longitudinal direction has never been
examined. The {\it Solar TErrestrial RElations Observatory} ({\it
STEREO}) mission started to observe CMEs in stereoscopic
view. Three-dimensional structure of the CMEs and their relation to
the associated flares should be tested again using {\it STEREO} data.

\acknowledgments

The authors would like to thank to the referee whose suggestions and
comments led to improvement of the manuscript. {\it SOHO} is a project
of international cooperation between ESA and NASA. The LASCO data used
here are produced by a consortium of the Naval Research Laboratory
(USA), Max-Planck-Institut fuer Aeronomie (Germany), Laboratoire
d'Astronomie (France), and the University of Birmingham (UK). Part of
this effort was supported by NASA (NNG05GR03G). Work done by G. M. was
partly supported by MNiSW through the grant N203 023 31/3055.


\end{document}